\documentclass[english,aps,prl,superscriptaddress,floatfix,notitlepage,reprint,show pacs]{revtex4-2}

\usepackage[T1]{fontenc}
\usepackage[utf8]{inputenc}
\usepackage{physics}
\usepackage{natbib}
\usepackage{amsthm}
\usepackage{float}
\usepackage{amssymb}
\usepackage{dsfont}
\usepackage{amsmath}
\usepackage{bm}
\usepackage{sublabel}
\usepackage{latexsym}
\usepackage{sidecap}
\usepackage{placeins}
\usepackage{url}
\makeatletter
\theoremstyle{plain}

\theoremstyle{plain}

\theoremstyle{plain}
\newtheorem*{prop*}{\protect\propositionname}

\usepackage{braket}
\usepackage{txfonts}
\usepackage{pifont}
\usepackage{graphicx}
\usepackage[usenames,dvipsnames]{xcolor}
\usepackage{hyperref}
\usepackage{cleveref}
\hypersetup{
    colorlinks=true,
    linkcolor=Red,       
    citecolor=blue,      
    urlcolor=cyan      
}
\usepackage{orcidlink}
\usepackage{tikz}
\usetikzlibrary{patterns,decorations.text,decorations.pathreplacing,decorations.pathmorphing}
\usepackage{caption}
\usepackage{subcaption}
\usepackage{rotating}
\usepackage{lipsum}
\usepackage{tocloft}

\date{\today}

\newcommand{\red}{\protect\tikz[baseline=-0.5ex]\draw[red] (0,0)--(0.5,0);}
\newcommand{\blue}{\protect\tikz[baseline=-0.5ex]\draw[blue] (0,0)--(0.5,0);}
\newcommand{\green}{\protect\tikz[baseline=-0.5ex]\draw[green] (0,0)--(0.5,0);}
\newcommand{\brown}{\protect\tikz[baseline=-0.5ex]\draw[brown] (0,0)--(0.5,0);}
\newcommand{\magenta}{\protect\tikz[baseline=-0.5ex]\draw[magenta] (0,0)--(0.5,0);}
\newcommand{\magentadashed}{\protect\tikz[baseline=-0.5ex]\draw[magenta,dashed] (0,0)--(0.5,0);}

\newcommand{\bluedashed}{\protect\tikz[baseline=-0.5ex]\draw[blue,dashed] (0,0)--(0.5,0);}

\newcommand{\orangedashed}{\protect\tikz[baseline=-0.5ex]\draw[orange] (0,0)--(0.5,0);}

\newcommand{\brownball}{\protect\tikz[baseline=-0.5ex]\draw[ball color=brown] (0,0) circle (0.1);}
\newcommand{\violet}{\protect\tikz[baseline=-0.5ex]\draw[ball color=violet] (0,0) circle (0.1);}
\begin{document}
\title{Relaxation Dynamics in Atomic-Molecular Bose Condensates in the Presence of Gaussian Noise}
\author{Avinaba Mukherjee \orcidlink{0009-0000-3765-6466}}
\thanks{\href{mailto:avinaba.mukherjee@rediffmail.com}{avinaba.mukherjee@rediffmail.com}}
\author{Raka Dasgupta \orcidlink{0000-0003-2148-4641}}
\thanks{\href{mailto:rdphy@caluniv.ac.in}{rdphy@caluniv.ac.in}}
\address{Department of Physics, University of Calcutta, $92$ A. P. C. Road, Kolkata $700009$, India}
\begin{abstract}
 We investigate the dynamics of atomic and molecular bosons weakly coupled via Feshbach detuning in the presence of Gaussian white noise. The time-evolution of the population imbalance between the two species, as well as the coherence of the system are analyzed using a Bloch sphere model. We observe that the system exhibits relaxation of the Bloch vector components towards a stable equilibrium. In the population imbalance dynamics,  the relaxation rates predicted by the Bogoliubov Born Green Kirkwood Yvon (BBGKY) hierarchy are found to be smaller than those calculated with a simple the mean field approximation. As for the coherence dynamics, the inclusion of correlations and fluctuations in the system can either increase or decrease the relaxation time, depending on the initial conditions. We attribute the increase in the relaxation time to the emergence of a structured noise, and the subsequent suppression of certain dephasing channels. We also study the impact of correlations and fluctuations on time-averaged quantities like the drift speed, the fringe visibility, the phase entanglement etc., and find the results to be in perfect agreement with the properties of the relaxation dynamics. 
 
\end{abstract}
\maketitle

\section{Introduction}

A system of coupled atomic-molecular Bose-Einstein condensates (BECs) acts as an interesting platform that has inspired a wide range of scientific investigations in recent times\cite{BEC1,BEC2,BEC3,BEC4,BEC5,BEC6,BEC7,BEC8,BEC9,BEC10,BEC11,BEC12,BEC13,BEC14,BEC15,BEC16}. Here, two atomic bosons can form a molecular boson via Feshbach resonance \cite{timmermans1999feshbach,kohler2006production}. This system effectively functions as a Bosonic Josephson Junction. Although conventional josephson junctions are created by trapping a BEC in a double well potential, \cite{BJJ1,BJJ2,BJJ3,BJJ4,BJJ5,BJJ6,BJJ7,BJJ8,BJJ9,BJJ10,BJJ11,BJJ12,BJJ13,BJJ14,BJJ15,BJJ16,BJJ17,BJJ18,BJJ19,BJJ20}, a coupled atomic-molecular BEC system serves a somewhat similar purpose because it, too, consists of two weakly
coupled macroscopic quantum states. The Feshbach coupling between the atomic and molecular bosons here plays the role of tunneling. In addition, the atomic-BEC (A-BEC) and molecular-BEC (M-BEC) states are separated by a threshold energy of molecule formation: analogous to the intermediate insulating layer in superconducting Josephson junctions. \\ 

The non-equilibrium dynamics of bosonic josephson junctions \cite{ref22,ref24,ref26,ref28,SBJJ,gati2007bosonic} is an emerging area of research. Two-mode BEC in a double well manifests a damped oscillation of population imbalance and relative phase towards a stable equilibrium point if the hopping amplitude and detuning are corrupted by Gaussian white Noise  \cite{stochastic_bosonic_josephson_junction}. If one considers quantum fluctuations described by the Bogoliubov Backreaction (BBR) formalism beyond the Mean-Field (MF) or classical regime in such a two-mode condensate, it is observed that decoherence in the system either enhances or suppresses phase diffusion \cite{Linblad_master_equation}. For an atomic-molecular BEC system, it was observed that the conversion dynamics in the presence of phase noise is well-described using MF for a short time scale, but for a longer time scale, the BBR correction is necessary \cite{bloch4}.\\

In this work, we theoretically study the relaxation dynamics of a Feshbach-coupled atomic-molecular BECs system, with a focus on the particle-imbalance between two quantum states, as well as the coherence. The system can be fully described in terms of the components of the Bloch vector: $\hat{L}_x$, $\hat{L}_y$ and $\hat{L}_z$. We find that when Gaussian White Noise corrupts the coupling strength and detuning, both MF and BBR treatment of the Hamiltonian lead to damped oscillatory dynamics for the Bloch vector components. However, the estimated relaxation time differs substantially from MF to BBR, i.e. it depends on higher-order correlation (self- or mutual) between the components. The longitudinal relaxation time $T_z$ and the transverse relaxation rate $T_{x-y}$ of the three components of the Bloch vector are substantially affected by the variance and covariance of $\hat{L}_x$, $\hat{L}_y$ and $\hat{L}_z$. We try to relate our results to several related physical parameters: (i) drift speed, (ii) fringe visibility, (iii) phase entanglement, and (iv) polar or precession angle of Bloch sphere.

 The divisions of the paper are as follows. In Sec. \ref{framework}, We establish the basic model, and derive the dynamical equations to include the effect of noise on the system. Then in Sec. \ref{Relaxation dynamics thermodynamic quantum}, we study the relaxation dynamics using both MF and BBR  frameworks. In Sec. \ref{long-trans}, we calculate the time-averaged values of quantities like fringe visibility, and phase entanglement, and connect these findings with the trends in the relaxation processes.
 In Sec. \ref{conclusion}, we summarize our results. 
 
\section{Basic Model}\label{framework}

We consider an ultracold system in which two bosonic atoms combine to form a bosonic molecule through a Feshbach resonance process\cite{AMBEC,AMBEC2,AMBEC3}. We treat the system in a two-channel model, consisting of a closed channel and an open or entrance channel \cite{review}. When the energy of the entrance channel matches that of the closed channel, the bosonic atoms are resonantly coupled to form bosonic dimers. The energy difference between the A-BEC and M-BEC states is the Feshbach detuning $\epsilon_b$, and this can be tuned using an external magnetic field. The system is described by the following Hamiltonian
\cite{similar_hamiltonian1,similar_hamiltonian2,similar_hamiltonian3,similar_hamiltonian4}.\\
\begin{equation}
\label{hamiltonian}
    \begin{split}
        \hat{H}= & \frac{u_1}{2} \hat{a}^\dagger \hat{a}^\dagger \hat{a}\hat{a}+ \frac{u_2}{2} \hat{b}^\dagger \hat{b}^\dagger \hat{b}\hat{b}+ u_3 \hat{a}^\dagger \hat{b}^\dagger \hat{b}\hat{a}\\ &+ g (\hat{a}^\dagger \hat{a}^\dagger \hat{b}+ \hat{b}^\dagger \hat{a}\hat{a}) + \epsilon_b  \hat{b}^\dagger \hat{b}
    \end{split}
\end{equation}
Here, $\hat{a}^\dagger$ and $\hat{a}$ are the atomic creation and annihilation operators, while $\hat{b}^\dagger$ and $\hat{b}$ are their molecular counterparts. Here, $u_1$ and $u_2$ represent atom-atom interaction and molecule-molecule interaction strength; $u_3$ is an interaction between bosonic atoms and bosonic molecules, and $g$ is the Feshbach coupling constant.\\

 In Sec. \ref{TSM}, we  recast the system Hamiltonian in terms of Bloch vector's components and then study the coherent evolution of each component. Next, Gaussian white noises are applied on both the Feshbach coupling strength and detuning,  and the incoherent evolution of the Bloch vector components are analyzed using the Ito calculus in Sec. \ref{ito}.
 
\subsection{Bloch vector description of the two-mode condensate}\label{TSM}

The Bloch vector's components or Schwinger pseudo-spin operators \cite{P70,P190} are introduced as follows \cite{bloch4, commutator}
\begin{subequations}
\label{operators}
    \begin{equation}
    \label{operators1}
        \hat{L}_x=\frac{\sqrt{2}}{N^\frac{3}{2}} (\hat{a}^\dagger \hat{a}^\dagger \hat{b} + \hat{b}^\dagger \hat{a} \hat{a})
    \end{equation}
    \begin{equation}
    \label{operators2}
        \hat{L}_y=\frac{\sqrt{2}i}{N^\frac{3}{2}} (\hat{a}^\dagger \hat{a}^\dagger \hat{b} - \hat{b}^\dagger \hat{a} \hat{a})
    \end{equation}
    \begin{equation}
    \label{operators3}
        \hat{L}_z=\frac{2\hat{b}^\dagger \hat{b}-\hat{a}^\dagger \hat{a}}{N}
    \end{equation}
    \begin{equation}
    \label{number operator}
      N=2\hat{b}^\dagger \hat{b}+\hat{a}^\dagger \hat{a}  
    \end{equation}
\end{subequations}

 $\hat{L}_x$ and $\hat{L}_y$  represent the real and imaginary parts of coherence, while $\hat{L}_z$ denotes the population imbalance between atoms and molecules \cite{cui2012atom}. $N$ is the total number of atoms in the system.\\

 In a semi-classical approximation, the operators $\hat{a}$ and $\hat{b}$ in Eq. (\ref{hamiltonian}) can be expressed as $\hat{a}=\sqrt{N_a}e^{i\theta_a}$, and $\hat{b}=\sqrt{N_b}e^{i\theta_b}$ \cite{similar_hamiltonian1,similar_hamiltonian3,motohashi2010particle,P272}. Here, $N_a$ represents the atomic population, and $N_b$ represents the molecular population where $\theta_a$, and $\theta_b$ are the phases of atomic and molecular condensate. 

\begin{subequations}
    \begin{equation}
    \label{non rigid1}
        \hat{L}_x=\frac{(1-\tilde{z})\sqrt{1+\tilde{z}}}{\sqrt{2}}cos\tilde{\phi}
    \end{equation}
    \begin{equation}
    \label{non rigid2}
        \hat{L}_y=\frac{(1-\tilde{z})\sqrt{1+\tilde{z}}}{\sqrt{2}}sin\tilde{\phi}
    \end{equation}
    \begin{equation}
    \label{non rigid3}
        \hat{L}_z=\tilde{z}
    \end{equation}
    \label{non rigid}
\end{subequations}

\begin{figure}[h]
\begin{center}
\includegraphics[width=0.8\linewidth]{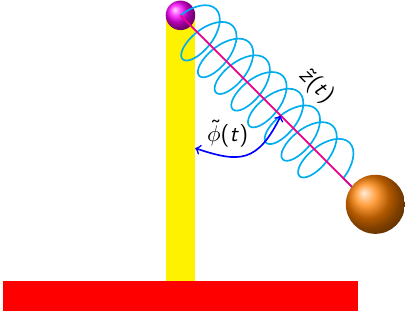}
\caption[short description]{The usual non-rigid pendulum model, where the point of suspension (\violet) is attached to the vertical stand. Since the pendulum is non-rigid, both the length $\tilde{z}(t)$ of the spring and the angle $\tilde{\phi}(t)$ of the bob (\brownball) are time-dependent variables.
}
\label{non_rigid_pendulum_model}
\end{center}
\end{figure}

\begin{figure}
\begin{center}
\includegraphics[width=0.7\linewidth]{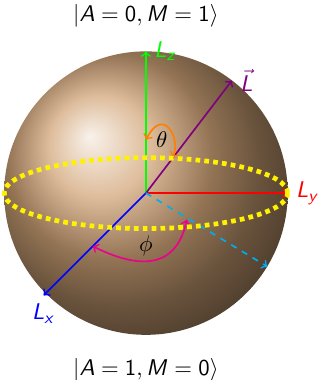}
\caption[short description]{The Bloch sphere model, where the North and South poles represent fully molecular (M) and atomic (A) Bose-Einstein condensates (BEC), respectively. The yellow circle denotes the equatorial plane.
}
\label{bloch sphere}
\end{center}
\end{figure}

Here, $z = 2\hat{b}^\dagger \hat{b}-\hat{a}^\dagger \hat{a} =2 N_b-N_a$ represents the difference between the number of atoms between the molecular state and the atomic state; and  $\tilde{\phi}=2\theta_a-\theta_b$  is the relative phase between the atomic-molecular BEC. Eq. (\ref{non rigid}) can be mapped to a non-rigid pendulum model as shown in Fig. (\ref{non_rigid_pendulum_model}). Here, $\tilde{z}$ and $\tilde{\phi}$ are canonically conjugate variables \cite{cui2012atom}.
We would like to point out that a Bloch vector model is often used to study the population dynamics of the BEC in a double well. For an atomic-molecular condensate, the Bloch vectors can be defined just in an analogous way. However, unlike the Bloch vectors for a single condensate system in a double well \cite{Linblad_master_equation,BEC10,split_correlation}, the components of the Bloch vector for the atomic-molecular condensate do not obey the usual SU(2) algebra \cite{bloch_vector4, fermion_number2, fermion_number4}. Still, they are very effective for the ease of visualization of the system dynamics. The fully molecular ($\ket{A=0,M=1}$) and fully atomic states ($\ket{A=1,M=0}$) can be depicted as the North and South poles \cite{ns_pole,P-33} of the Bloch Sphere in Fig. (\ref{bloch sphere}).

 The commutation relations of the Bloch vector components are given below \cite{bloch4,liu2010shapiro}
\begin{subequations}
    \begin{equation}
    \label{commutation1}
        [\hat{L}_x , \hat{L}_y]=\frac{i}{N}(1-\hat{L}_z)(1+3\hat{L}_z)+O(\frac{1}{N^2})
    \end{equation}
    \begin{equation}
    \label{commutation2}
        [\hat{L}_y , \hat{L}_z]= \frac{4i\hat{L}_x}{N}
    \end{equation}
    \begin{equation}
    \label{commutation3}
        [\hat{L}_z , \hat{L}_x]= \frac{4i\hat{L}_y}{N}
    \end{equation}
     \label{commutation}
\end{subequations}

Moreover,  $[N , \hat{L}_i]=0$ for $i\in{x,y,z}$.

 Now, Eq. (\ref{hamiltonian}) recast in terms of Bloch vector's components becomes
\begin{equation}
\label{hamiltonian bloch}
    \begin{split}
        \hat{H}= & \frac{u_1 N^2}{8}(\hat{L}_z-1)^2+\frac{u_1N}{4}(\hat{L}_z-1)+\frac{u_2N^2}{32}(\hat{L}_z+1)^2\\&-\frac{u_2N}{8}(\hat{L}_z+1)-\frac{u_3N^2}{8}(\hat{L}^2_z-1)+\frac{gN^{\frac{3}{2}}}{\sqrt{2}}\hat{L}_x+\frac{N\epsilon_b}{4}(\hat{L}_z+1)
    \end{split}
\end{equation}
Rewriting  $N{u_1}=U_1$, $N{u_2}=U_2$, $N{u_3}=U_3$, and $g\sqrt{N}=\tilde{g}$.  Eq. (\ref{hamiltonian bloch}) can be  expressed as 
\begin{equation}
\label{hamiltonian table}
    \begin{split}
        \hat{H}= & \frac{U_1N}{8}(\hat{L}_z-1)^2+\frac{U_1}{4}(\hat{L}_z-1)+\frac{U_2N}{32}(\hat{L}_z+1)^2\\&-\frac{U_2}{8}(\hat{L}_z+1)-\frac{U_3N}{8}(\hat{L}^2_z-1)+\frac{\tilde{g}N}{\sqrt{2}}\hat{L}_x+\frac{N\epsilon_b}{4}(\hat{L}_z+1)
    \end{split}
\end{equation}

In usual experiments, \cite{large_N1, large_N2} the number of particles can be taken as ~$10^6$-$10^7$. Now, in large N limit, Eq. (\ref{hamiltonian table}) becomes,
 \begin{equation}
\label{hamiltonian large N}
    \begin{split}
        \hat{\mathcal{H}}=\frac{\hat{H}}{N}= & \frac{U_1}{8}(\hat{L}_z-1)^2+\frac{U_2}{32}(\hat{L}_z+1)^2\\&-\frac{U_3}{8}(\hat{L}^2_z-1)+\frac{\tilde{g}}{\sqrt{2}}\hat{L}_x+\frac{\epsilon_b}{4}(\hat{L}_z+1)
    \end{split}
\end{equation}
The dynamical evolution of the Bloch vector's components, following Heisenberg's equation of motion, is given by :
\begin{subequations}
    \begin{equation}
    \begin{split}
        \dot{\hat{L}}_x= & \Bigg(\frac{U_3}{2\hbar}-\frac{U_2}{8\hbar}-\frac{U_1}{2\hbar}\Bigg)(\hat{L}_y \hat{L}_z + \hat{L}_z \hat{L}_y)  \\ &+ \Bigg(\frac{U_1}{\hbar}-\frac{U_2}{4\hbar}-\frac{\epsilon_b}{\hbar}\Bigg) \hat{L}_y
        \end{split}
    \end{equation}
    \begin{equation}
    \begin{split}
        \dot{\hat{L}}_y= & \Bigg(\frac{U_1}{2\hbar}+\frac{U_2}{8\hbar}-\frac{U_3}{2 \hbar}\Bigg) \Big(\hat{L}_x \hat{L}_z+ \hat{L}_z \hat{L}_x\Big) \\ & + \Bigg(\frac{U_2}{4\hbar}-\frac{U_1}{\hbar} +\frac{\epsilon_b}{\hbar}\Bigg)\hat{L}_x\\ &-\frac{\tilde{g}}{\sqrt{2}\hbar} \{(1-\hat{L}_z)(1+3\hat{L}_z)\}
        \end{split}
    \end{equation}
    \begin{equation}
       \dot{\hat{L}}_z=\frac{2\sqrt{2}\tilde{g}}{\hbar}\hat{L}_y 
    \end{equation}
   \end{subequations}

It is also found that $\dot{N}=0$, i.e., the total number of atoms remain conserved.

In the following subsection (Sec.\ref{ito}), we investigate the effect of applying Gaussian white noise to the parameters $\tilde{g}$ and $\epsilon_b$. This noise drives the system out of equilibrium, and we analyze how the system relaxes back towards its equilibrium  configuration.

\subsection{Inclusion of Noise in the dynamical equations}\label{ito}
Next, we consider that the modified coupling ($\tilde{g}$) and the detuning ($\epsilon_b$) are corrupted by white noise $n_x(t)$ and $n_z(t)$, respectively. These noises are delta-correlated Gaussian noises and their properties are as follows \cite{noise_property} 
\begin{subequations}
    \begin{equation}
    \langle n_x (t) \rangle =0
    \end{equation}
    \begin{equation}
    \langle n_z (t) \rangle =0 
    \end{equation}
     \begin{equation}
    \langle n_x (t_1) n_x (t_2) \rangle =\Gamma_x \delta (t_1-t_2)
    \end{equation}
    \begin{equation}
    \langle n_z (t_1) n_z (t_2) \rangle =\Gamma_z \delta (t_1-t_2)
    \end{equation}
    \end{subequations}
Here $\Gamma_x$ and $\Gamma_z$ are the strength of the noises applied to $\tilde{g}$ and $\epsilon_b$ respectively\cite{P225}. The physical origin of $\Gamma_x$ is the decoherence or dephasing due to elastic collisions between the thermal cloud and condensed particles \cite{witthaut2008dissipation, anglin1997cold,ruostekoski1998bose}. On the other hand, the physical origin of $\Gamma_z$ is the fluctuation of the applied magnetic field around the Feshbach resonance point \cite{bloch_vector4} or the thermal fluctuation of the BEC at finite temperatures \cite{saha2023phase}.

Accordingly, the dynamical relations of the Bloch vector's components are expressed as follows:   
\begin{subequations}
    \begin{equation}
    \begin{split}
        \dot{\hat{L}}_x=& \Bigg(\frac{U_3}{2\hbar}-\frac{U_2}{8\hbar}-\frac{U_1}{2\hbar}\Bigg)(\hat{L}_y\hat{L}_z+\hat{L}_z\hat{L}_y)\\&+\Bigg(\frac{U_1}{\hbar}-\frac{U_2}{4\hbar}-\frac{\epsilon_b+n_z(t)}{\hbar}\Bigg)\hat{L}_y
        \end{split}
    \end{equation}
    \begin{equation}
    \begin{split}
        \dot{\hat{L}}_y=& \Bigg(\frac{U_1}{2\hbar}+\frac{U_2}{8\hbar}-\frac{U_3}{2\hbar}\Bigg)(\hat{L}_x\hat{L}_z+\hat{L}_z\hat{L}_x)\\&+\Bigg(\frac{U_2}{4\hbar}-\frac{U_1}{\hbar}+\frac{\epsilon_b+n_z(t)}{\hbar}\Bigg)\hat{L}_x\\&-\frac{\tilde{g}+n_x(t)}{\sqrt{2}\hbar}(1-\hat{L}_z)(1+3\hat{L}_z)
        \end{split}
    \end{equation}
    \begin{equation}
       \dot{\hat{L}}_z=\frac{2\sqrt{2}\bigg(\tilde{g}+n_x(t)\bigg)}{\hbar}\hat{L}_y 
    \end{equation}
\end{subequations}

Here $n_x={dw_x}/{dt}$ and $n_z={dw_z}/{dt}$  \cite{noise_property,noise_property2} where $w_x$ and $w_z$ are Wiener processes.
\begin{subequations}
    \begin{equation}
    \begin{split}
        d\hat{L}_x= &\Bigg[ \Bigg(\frac{U_3 }{2 \hbar}-\frac{U_2 }{8\hbar}-\frac{U_1 }{2\hbar}\Bigg) (\hat{L}_y \hat{L}_z + \hat{L}_z \hat{L}_y)  \\ &+ \Bigg(\frac{U_1 }{\hbar}-\frac{U_2 }{4\hbar}-\frac{\epsilon_b}{\hbar}\Bigg) \hat{L}_y\Bigg]dt-\frac{\hat{L}_y}{\hbar}dw_z
        \end{split}
    \end{equation}
    \begin{equation}
    \begin{split}
        d{\hat{L}_y}= &\Bigg[ \Bigg(\frac{U_1 }{2 \hbar}+\frac{U_2 }{8\hbar}-\frac{U_3 }{2\hbar}\Bigg) (\hat{L}_x \hat{L}_z+ \hat{L}_z \hat{L}_x) \\ & +\Bigg(\frac{U_2 }{4\hbar}-\frac{U_1 }{\hbar} +\frac{\epsilon_b}{\hbar}\Bigg)\hat{L}_x-\frac{\tilde{g}}{\sqrt{2}\hbar}  (1-\hat{L}_z)(1+3\hat{L}_z)\Bigg]dt\\& +\frac{\hat{L}_x}{\hbar}dw_z -\frac{(1-\hat{L}_z)(1+3\hat{L}_z)}{\sqrt{2}\hbar}dw_x
        \end{split}
    \end{equation}
    \begin{equation}
       d{\hat{L}_z}=\frac{2\sqrt{2}\tilde{g}}{\hbar}\hat{L}_ydt +\frac{2\sqrt{2}\hat{L}_y}{\hbar}dw_x
    \end{equation}
\end{subequations}
We have used Ito calculus on $dL_j$ \cite{noise_property2}. Note that  $\langle{dw_{x}}\rangle=\langle{dw_{z}}\rangle=0$, $dw_xdw_x=\gamma_x dt$ and $dw_z dw_z=\gamma_z dt$ \cite{noise_property,wiener,noise_property2} where $\gamma_x$ and $\gamma_z$ are the strengths of Wiener processes.
\begin{subequations}
    \begin{equation}
    \begin{split}
        d\hat{L}_x= &\Bigg[ \Bigg(\frac{U_3 }{2\hbar}- \frac{U_2 }{8\hbar}-\frac{U_1 }{2\hbar}\Bigg) (\hat{L}_y \hat{L}_z + \hat{L}_z \hat{L}_y)  \\ &+ \Bigg(\frac{U_1 }{\hbar}-\frac{U_2 }{4\hbar}-\frac{\epsilon_b}{\hbar}\Bigg) \hat{L}_y-\frac{\gamma_z \hat{L}_x}{2\hbar^2}\Bigg]dt-\frac{\hat{L}_y}{\hbar}dw_z
        \end{split}
    \end{equation}
    \begin{equation}
    \begin{split}
        d{\hat{L}_y}= &\Bigg[ \Bigg(\frac{U_1 }{2\hbar}+\frac{U_2 }{8\hbar}-\frac{U_3 }{2 \hbar}\Bigg) (\hat{L}_x \hat{L}_z+ \hat{L}_z \hat{L}_x) \\ & +\Bigg(\frac{U_2 }{4\hbar}-\frac{U_1 }{\hbar} +\frac{\epsilon_b}{\hbar}\Bigg)\hat{L}_x-\frac{\tilde{g}}{\sqrt{2}\hbar}  (1-\hat{L}_z)(1+3\hat{L}_z)\\ &+\frac{1}{2}\{-\frac{4}{\hbar^2}\hat{L}_y+\frac{6}{\hbar^2}(\hat{L}_y\hat{L}_z-\hat{L}_z\hat{L}_y)\}\gamma_x-\frac{\hat{L}_y\gamma_z}{2\hbar^2}]dt \\&-\frac{(1-\hat{L}_z)(1+3\hat{L}_z)}{\sqrt{2}\hbar}dw_x+\frac{\hat{L}_x}{\hbar}dw_z
        \end{split}
    \end{equation}
    \begin{equation}
       d{\hat{L}_z}=\Bigg(\frac{2\sqrt{2}\tilde{g}\hat{L}_y}{\hbar}-\frac{\gamma_x(1-\hat{L}_z)(1+3\hat{L}_z)}{\hbar^2}\Bigg)dt +\frac{2\sqrt{2}\hat{L}_y}{\hbar}dw_x
    \end{equation}
\end{subequations}
If an average is taken over the noises, and the operators are replaced by their expectation values, then one obtains
\begin{subequations}
\label{expectations}
    \begin{equation}
    \label{expectation1}
    \begin{split}
       \frac{d\langle \hat{L}_x\rangle}{dt}= & \Bigg( \frac{U_3 }{2\hbar}-\frac{U_2 }{8\hbar}-\frac{U_1 }{2\hbar}\Bigg) \langle \hat{L}_y \hat{L}_z + \hat{L}_z \hat{L}_y\rangle  \\ &+ \Bigg(\frac{U_1 }{\hbar}-\frac{U_2 }{4\hbar}-\frac{\epsilon_b}{\hbar}\Bigg)\langle \hat{L}_y\rangle-\frac{\gamma_z \langle \hat{L}_x\rangle}{2\hbar^2} 
        \end{split}
    \end{equation}
    \begin{equation}
    \label{expectation2}
    \begin{split}
        \frac{d\langle \hat{L}_y \rangle}{dt}= & \left(\frac{U_1 }{2\hbar}+\frac{U_2 }{8\hbar}-\frac{U_3 }{2 \hbar}\right) \langle \hat{L}_x \hat{L}_z+ \hat{L}_z \hat{L}_x\rangle \\ & +\left(\frac{U_2 }{4\hbar}-\frac{U_1 }{\hbar} +\frac{\epsilon_b}{\hbar}\right)\langle \hat{L}_x\rangle-\frac{\tilde{g}}{\sqrt{2}\hbar}  \langle (1+2\hat{L}_z-3\hat{L}^2_z)\rangle \\ &+\frac{\gamma_x}{\hbar^2}\langle (3\hat{L}_y\hat{L}_z+3\hat{L}_z\hat{L}_y-2\hat{L}_y)\rangle-\frac{\gamma_z \langle \hat{L}_y\rangle }{2\hbar^2}
        \end{split}
    \end{equation}
    \begin{equation}
    \label{expectation3}
    \begin{split}
      \frac{d\langle \hat{L}_z\rangle}{dt}=\frac{2\sqrt{2}\tilde{g}\langle \hat{L}_y\rangle }{\hbar}-\frac{\gamma_x\langle (1+2\hat{L}_z-3\hat{L}^2_z)\rangle}{\hbar^2}
       \end{split}
    \end{equation}
\end{subequations}

In the next section (Sec. \ref{Relaxation dynamics thermodynamic quantum}), We examine the relaxation dynamics of the Bloch vector components in both the thermodynamic (MF) and quantum (BBGKY) limits.

\section{Relaxation Dynamics}\label {Relaxation dynamics thermodynamic quantum}

We study the relaxation dynamics of the Bloch vector components. First, we formulate the dynamical equations by considering only the first moment of the Bloch vector components using the mean-field (MF) approximation. Next, we extend that calculation using the Bogoliubov-Born-Green-Kirkwood-Yvon (BBGKY) framework, considering nonzero variance and covariance of these components. We solve these equations numerically for both the MF (Sec. \ref{MF}) and the BBGKY calculations (Sec. \ref{BBR}). 

\subsection{MF}{\label{MF}}
\begin{figure}
\centering
\begin{subfigure}{0.8\linewidth}
    \centering
    \includegraphics[width=\linewidth]{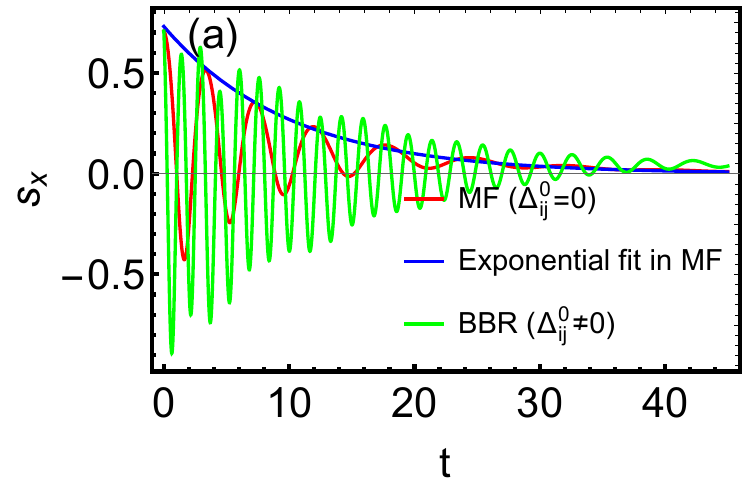}
    \phantomcaption 
    \label{relaxation s_a}
\end{subfigure}
\begin{subfigure}{0.8\linewidth}
    \centering
    \includegraphics[width=\linewidth]{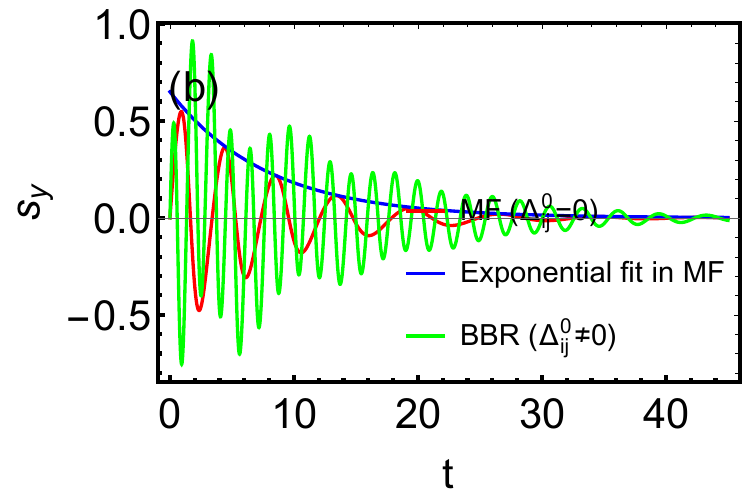}
    \phantomcaption 
    \label{relaxation s_b}
\end{subfigure}
\begin{subfigure}{0.8\linewidth}
    \centering
    \includegraphics[width=\linewidth]{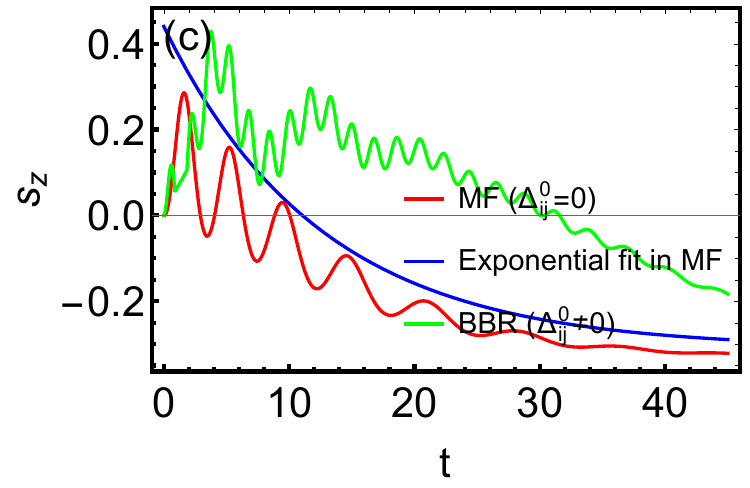}
    \phantomcaption 
    \label{relaxation s_c}
\end{subfigure}
\caption[short description]{Mean-field (MF) ($\Delta^0_{ij} = 0$), beyond mean-field (BBR) ($\Delta^0_{ij} \neq 0$), and the analytical expression of the relaxation rate in MF are denoted by the (\red), (\green), and (\blue) curves, respectively, where $\Delta^0_{ij}$ represents the correlation at $t=0$.  

Additionally, the relaxation dynamics of (a) $s_x$, (b) $s_y$, and (c) $s_z$ are considered.
}
\label{relaxation s}
\end{figure}
In the mean-field framework, the system is described using the expectation values of the Bloch vectors, and all higher-order correlations are neglected. Denoting $\langle \hat{L}_i\rangle$ by $s_i$ for $i\in x,y,z$, we derive a set of coupled nonlinear equations in terms of $s_x$, $s_y$, and $s_z$.

\subsubsection{Mathematical Framework}
Under the mean-field approximation, a two-point correlator involving the Bloch vectors can be factorized as the product of their individual expectation values. Thus, $\langle \hat{L}_x \hat{L}_z\rangle=\langle \hat{L}_z \hat{L}_x\rangle=\langle \hat{L}_x\rangle\langle \hat{L}_z\rangle$ and $\langle \hat{L}_y \hat{L}_z\rangle=\langle \hat{L}_z \hat{L}_y\rangle=\langle \hat{L}_y\rangle\langle \hat{L}_z\rangle$ \cite{Linblad_master_equation}. Writing $c_1={U_3}/{2\hbar}-{U_1}/{2\hbar}-{U_2}/{8\hbar}$ and $c_2=U_1/\hbar-{U_2}/{4\hbar}-\epsilon_b/\hbar$, we arrive at the Bloch equations similar to the well-known NMR process \cite{bloch1946nuclear,viola2000stochastic}.
\begin{subequations}
\label{final}
    \begin{equation}
    \label{final1}
    \begin{split}
       \dot{s_x}=2c_1s_ys_z+c_2s_y-\frac{\gamma_zs_x}{2} 
        \end{split}
    \end{equation}
    \begin{equation}
    \label{final2}
    \begin{split}
        \dot{s_y}=&-2c_1s_xs_z-c_2s_x-\frac{\tilde{g}}{\sqrt{2}}(1+2s_z-3s^2_z)\\&+\gamma_x(-2s_y+6s_ys_z)-\frac{\gamma_z s_y}{2}
        \end{split}
    \end{equation}
    \begin{equation}
    \label{final3}
    \begin{split}
      \dot{s_z}=2\sqrt{2}\tilde{g}s_y-\gamma_x(1+2s_z-3s^2_z)
       \end{split}
    \end{equation}
\end{subequations}

Eqs. (\ref{final1}), (\ref{final2}), and (\ref{final3}) can be numerically solved for a given set of initial conditions. To obtain realistic values of the parameters $c_1$, $c_2$ and $\tilde{g}$, we recall that in Feshbach resonance experiments with ultracold atoms $u_1={4\pi \hbar^2 a_{\text{bg}}}/{m_{atom}}$, $u_2={4\pi \hbar^2 a_{bb}}/{m_{\text{molecule}}}$, $u_3={4\pi \hbar^2 a_{ab}}/{\mu}$, $g=\sqrt{\mu_{co}\Delta B u_1}$, $\mu=m_a m_b/(m_a+m_b)$, and $\epsilon_b=\mu_{\text{co}}(B-B_0)$. Here, $a_{\text{bg}}$ and $a_{\text{bb}}$ are the background scattering lengths of bosonic atoms and bosonic molecules, and $a_{\text{ab}}$ is the scattering length corresponding to atom-molecule interaction. Also, $m_{\text{atom}}$, $m_{\text{molecule}}$ and $\mu$ are the mass of a bosonic atom, the mass of a bosonic molecule and the reduced mass of a combined bosonic atom-molecule structure respectively. $\mu_{co}$ and $\Delta B$ represent the magnetic moment and resonance width of the bosonic atom, respectively, where $B$ is the applied magnetic field and $B_0$ is the resonance position

\subsubsection{Dynamics of the Bloch Vectors}{\label{initial condition}}

For numerical solutions, we use the parameters corresponding to Feshbach resonance of ${}^{87}\mathrm{Rb}$. Note that $\Delta B$ and $B_0$ have the experimental values $0.21$ Gauss and $1007.4$ Gauss, respectively \cite{kohler2006production} for this system. Here, we take $\lvert B-B_0\rvert=10$ Gauss. In this limit, the values of $U_1$, $U_2$, $U_3$, $\tilde{g}$ and $\epsilon_b$ are $1$, $2$, $-1.5$, $0.2$, and $0.191$, respectively \cite{strecker2003conversion,jin2005quantum,liu2008role,kohler2006production}. We choose $\gamma_x$ and $\gamma_z$ to be $10\%$ of $\tilde{g}$ and $\epsilon_b$, respectively. So, the system parameters become: $c_1=c_2=-1.5$, $\gamma_x=0.02$, $\gamma_z=0.2$. Using Eq. \ref{non rigid}, we construct our initial values for the Bloch vector components: $s^0_x=0.707$, $s^0_y=0$, and $s^0_z=0$ (essentially, $\tilde{z}=0$ and $\tilde{\phi}=0$). Thus, our starting point is the Josephson state $\mathbf{0}$ \cite{marino1999bose,saha2023phase} which corresponds to $\tilde{\phi}=0$.

\begin{figure}
    \centering
    \includegraphics[width=0.9\linewidth]{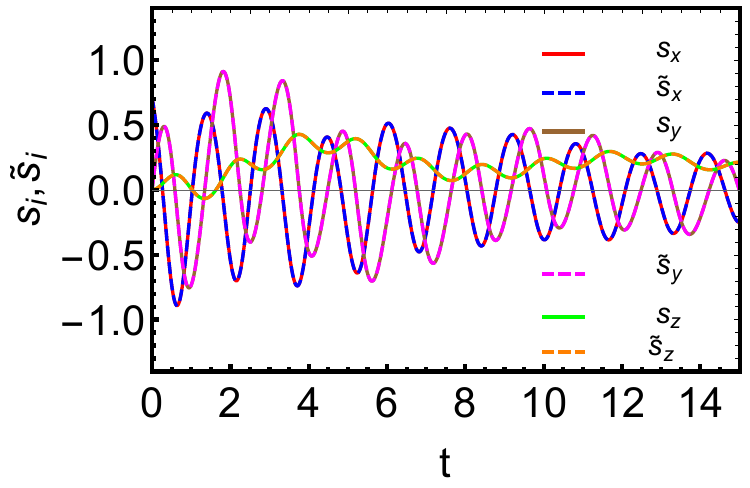}
\caption[Short description]{ The Bloch vector components show similar dynamical behavior whether the terms $2s_i s_j$ or $s_i s_j + s_j s_i$ are incorporated into the equations, denoted by $s_i$ and $\tilde{s}_i$, respectively. The components are represented as follows:  
$s^{2s_i s_j}_x$ as $s_x$ (\red),  
$s^{s_i s_j+s_js_i}_x$ as $\tilde{s}_{x}$ (\bluedashed),  
$s^{2s_i s_j}_y$ as $s_y$ (\brown),  
$s^{s_i s_j+s_js_i}_y$ as $\tilde{s}_{y}$ (\magenta),  
$s^{2s_i s_j}_z$ as $s_z$(\green),  
$s^{s_i s_j+s+js_i}_z$ as $\tilde{s}_{z}$(\orangedashed).  
Each component is represented by its respective color.}
\label{large N limit}
\end{figure}

\begin{figure}
    \centering
    \includegraphics[width=0.9\linewidth]{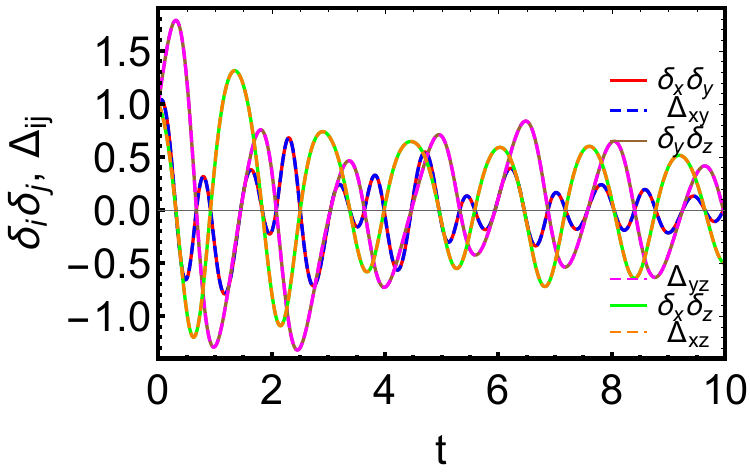}
\caption[Short description]{The quantities $\delta_i$, $\delta_j$, and $\Delta_{ij}$ exhibit the same dynamics. Their representations are as follows:  
$\delta_x \delta_y$ (\red),  
$\Delta_{xy}$ (\bluedashed),  
$\delta_y \delta_z$ (\brown),  
$\Delta_{yz}$ (\magentadashed),  
$\delta_x \delta_z$ (\green),  
$\Delta_{xz}$ (\orangedashed).  
Each quantity is denoted by its respective color.
}
\label{Delta delta}
\end{figure}

MF oscillations are plotted using the red-damped curve in Figs. (\ref{relaxation s_a}), (\ref{relaxation s_b}), and (\ref{relaxation s_c}) where $\Delta^0_{ij}$ stands for $\Delta_{ij}$ at $t=0$. From Eq. (\ref{final}) we obtain the relaxation rate of $s_x$, $s_y$, and $s_z$. For this, we linearize the Eq. (\ref{final3}) about the stable equilibrium point $(0,0,-1/3)$. Now, the relaxation rates are the following,
\begin{subequations}
\label{relaxation rate}
    \begin{equation}
    \label{relaxation rate1}
    \begin{split}
    \frac{1}{T^{\text{MF}}_x}=\frac{\gamma_z}{2}
         \end{split}
    \end{equation}
    \begin{equation}
    \label{relaxation rate2}
    \begin{split}
    \frac{1}{T^\text{MF}_y}=2\gamma_x (1-3s_z)+\frac{\gamma_z}{2}
        \end{split}
    \end{equation}
    \begin{equation}
    \label{relaxation rate3}
    \begin{split}
    \frac{1}{T^\text{MF}_z}=4\gamma_x
       \end{split}
    \end{equation}
\end{subequations}

The blue curves in Figs. (\ref{relaxation s_a}), (\ref{relaxation s_b}), and (\ref{relaxation s_c}) represent the exponential fits of the damping process in the mean field (MF) approximation. Here, we have modeled $s_x$, $s_y$, and $s_z$ as $\approx A_xe^{-t/T_x}$, $\approx A_ye^{-t/T_y}$, and $\approx A_{z1}e^{-t/T_z}-A_{z2}$ against $t$ as obtained from Eq. (\ref{relaxation rate}). It should be noted that the peaks of the MF curve (red) fall on this exponential fit (blue) for each $s_i$. Here, $A_x$, $A_y$, and $A_{z1}$ are the exponential fit amplitudes for $s_x$, $s_y$, and $s_z$, respectively. There is a small downward shift $A_{z2}$ in the $s_z$ vs. $t$ plot to capture particle leakage from the M-BEC state to the A-BEC state. For this set of graphs, $A_x=0.73$, $A_y=0.65$, $A_{z1}=0.75$, and $A_{z2}=0.31$.

In the next subsection (Sec. \ref{BBR}), we discuss the BBGKY or BBR method that successfully incorporates the second moments (variance and covariance) of the bloch vector components in the dynamics. For a quantum system, both the variance and the covariance are non-trivial,  and play an important role in the relaxation process.

\subsection{BBR} {\label {BBR}}

The mean-field (MF) results are insufficient for accurately describing strongly correlated quantum systems. Therefore, we need to incorporate higher-order expectation values in the dynamical equations. In this section, we present a modified MF theory for the atomic-molecular BEC, by employing up to the second order of the Bogoliubov-Born-Green-Kirkwood-Yvon (BBGKY) hierarchy in the equations of motions. This truncation scheme is often termed the Bogoliubov Backreaction (BBR) method as well. 

\subsubsection{Mathematical Framework}
 Here, we decompose the expectations of three operators as follows.\\
\begin{equation}
\begin{split}
   \label{higherarchy}
    \langle \hat{L}_i\hat{L}_j\hat{L}_k\rangle&\approx \langle \hat{L}_i\hat{L}_j\rangle \langle \hat{L}_k\rangle+\langle \hat{L}_i\rangle+\langle \hat{L}_j\hat{L}_k\rangle+\langle \hat{L}_i\hat{L}_k\rangle \langle \hat{L}_j\rangle\\&-2\langle \hat{L}_i\rangle \langle \hat{L}_j\rangle \langle \hat{L}_k\rangle 
\end{split}
\end{equation}

Moreover, we define the two-point correlation function $\Delta_{ij}$ as\\
\begin{equation}
    \label{normal correlation}
    \Delta_{ij}=\langle \hat{L}_i\hat{L}_j+\hat{L}_j\hat{L}_i\rangle-2\langle \hat{L}_i\rangle \langle \hat{L}_j\rangle
\end{equation}

If $\Delta_{ij}\neq 0$, we enter the BBR regime; otherwise, we stay in the MF regime where $\Delta_{ij}=0$. In other words, $\Delta_{ij}=0$ implies that the system remains in the eigenstates of the observable \cite{P-33,P-5}. 
 In terms of 1st and 2nd order moments, Eq. (\ref{expectations}) becomes
\begin{subequations}
\label{covariance}
    \begin{equation}
    \label{covariance1}
    \begin{split}
    \dot{s_x}=c_1(\Delta_{yz}+2s_ys_z)+c_2s_y-\frac{\gamma_z s_x}{2}
         \end{split}
    \end{equation}
    \begin{equation}
    \label{covariance2}
\begin{split}
    \dot{s_y}=&-c_1(\Delta_{zx}+2s_zs_x)-c_2s_x-\frac{\tilde{g}}{\sqrt{2}}(1+2s_z-\frac{3}{2}\Delta_{zz}-3s^2_z)\\&+\gamma_x(-2s_y+3\Delta_{yz}+6s_ys_z)-\frac{\gamma_z s_y}{2}
        \end{split}
    \end{equation}
    \begin{equation}
    \label{covariance3}
    \begin{split}
    \dot{s_z}=2\sqrt{2}\tilde{g} s_y-\gamma_x(1+2s_z-\frac{3}{2}\Delta_{zz}-3s^2_z)
       \end{split}
    \end{equation}
    \begin{equation}
    \label{covariance4}
\begin{split}
    \dot{\Delta}_{xx}=4c_1s_y\Delta_{zx}+2(2c_1s_z+c_2)\Delta_{xy}-\gamma_z\Delta_{xx}
        \end{split}
        \end{equation}
        \begin{equation}
    \label{covariance5}
\begin{split}
    \dot{\Delta}_{yy}=&\bigg(-4c_1s_x-2\sqrt{2}\tilde{g} (1-3s_z)+12\gamma_xs_y\bigg)\Delta_{yz}\\&-2(2c_1s_z+c_2)\Delta_{xy}-\bigg(\gamma_z+4\gamma_x(1-3s_z)\bigg)\Delta_{yy}
        \end{split}
        \end{equation}
        \begin{equation}
    \label{covariance6}
\begin{split}
    \dot{\Delta}_{zz}=4\sqrt{2}\tilde{g}\Delta_{yz}-4\gamma_x (1-3s_z)\Delta_{zz}
        \end{split}
        \end{equation}
        \begin{equation}
    \label{covariance7}
\begin{split}
    \dot{\Delta}_{xy}=&-(2c_1s_z+c_2)\Delta_{xx}+(2c_1s_z+c_2)\Delta_{yy}+2c_1s_y\Delta_{yz}\\&+\bigg(6\gamma_xs_y-2c_1s_x-\sqrt{2}\tilde{g}(1-3s_z)\bigg)\Delta_{zx}\\&-\bigg(\gamma_z+2\gamma_x(1-3s_z)\bigg)\Delta_{xy}
        \end{split}
        \end{equation}
        \begin{equation}
    \label{covariance9}
\begin{split}
    \dot{\Delta}_{yz}=&\bigg(-2c_1s_x-\sqrt{2}\tilde{g}(1-3s_z)+6\gamma_x s_y\bigg)\Delta_{zz}-(2c_1s_z+c_2)\Delta_{zx}\\&+2\sqrt{2}\tilde{g}\Delta_{yy}-\bigg(\frac{\gamma_z}{2}+4\gamma_x(1-3s_z)\bigg)\Delta_{yz}
        \end{split}
        \end{equation}
  \begin{equation}
    \label{covariance8}
\begin{split}
    \dot{\Delta}_{zx}=&2c_1s_y\Delta_{zz}+(2c_1s_z+c_2)\Delta_{yz}+2\sqrt{2}\tilde{g}\Delta_{xy}\\&-\bigg(2\gamma_x(1-3s_z)+\frac{\gamma_z}{2}\bigg)\Delta_{zx}
        \end{split}
        \end{equation}
        
\end{subequations}

\subsubsection{Dynamics of the Bloch vectors}{\label {sxyzBBR}}

In addition to the initial values specified in Sec. \ref{initial condition}, we consider variances and covariances $\Delta^0_{ii}=\Delta^0_{ij}=1$ to study the dynamics of $s_x$, $s_y$ and $s_z$ in Figs. (\ref{relaxation s_a}), (\ref{relaxation s_b}), and (\ref{relaxation s_c}). The green damped oscillatory curves here represent the BBR solutions. It is evident that in the BBR approximation, all three components ($s_x$, $s_y$, and $s_z$) have a longer relaxation time compared to their respective mean-field (MF) counterparts, indicating that a fully correlated system requires more time to relax. However, when certain individual fluctuations ($\Delta^0_{ii}$) or correlations ($\Delta^0_{ij}$) are present, they can lead to longer or shorter relaxation times compared to the MF values, depending on the individual initial conditions.  

Additionally,  we make two interesting observations. First, if terms like $(s_is_j+s_js_i)$ in Eq. (\ref{covariance}) are replaced by $2s_is_j$, the dynamics remains the same. This is demonstrated in Fig. (\ref{large N limit}), where $\tilde{s}_x$, $\tilde{s}_y$, $\tilde{s}_z$ are solutions of the original equations; and $s_x$, $s_y$ and $s_z$ are solutions of the same set of equations but with $(s_is_j+s_js_i)$ substituted by $2s_is_j$. This shows that we are in the quasi-classical regime, which is consistent with our initial assumption of a large $N$ limit. This is similar to the situation that arises for BEC in a double well \cite{graefe2010classical,graefe2010quantum,graefe2008mean,pi2024dynamics,soares2023dominant,graefe2013mean}.

Secondly, the terms $\Delta_{ij}$ can be written as a product of $\delta_i$ and $\delta_j$ throughout the course of evolution. This can be illustrated by splitting $\Delta^0_{ij}$ as $\delta^0_i\delta^0_j$ for each $i$ and $j$  and noting that $\Delta_{ij}(t)$ remains the same as $\delta_i(t)\delta_j(t)$  for all subsequent times. The dynamics of $\Delta_{xy}$, $\delta_x\delta_y$, $\Delta_{yz}$, $\delta_y\delta_z$, $\Delta_{xz}$, and $\delta_x\delta_z$ are represented in Fig.~(\ref{Delta delta}). This, again, resembles the situation with two-mode BEC in a double well \cite{split_correlation}. The dynamics of $\delta_x$, $\delta_y$, and $\delta_z$ here are given by 
\begin{subequations}
\label{perturbation}
    \begin{equation}
    \label{perturbation1}
       \dot{\delta}_x=2c_1s_y\delta_z+(2c_1s_z+c_2)\delta_y-\frac{\gamma_z}{2}\delta_x
    \end{equation}
    \begin{equation}
    \label{perturbation2}
    \begin{split}
     \dot{\delta}_y=&\bigg(-2c_1s_x-\sqrt{2}\tilde{g}(1-3s_z)+6\gamma_xs_y\bigg)\delta_z-(2c_1s_z+c_2)\delta_x\\&-\bigg(2\gamma_x(1-3s_z)+\frac{\gamma_z}{2}\bigg)\delta_y     \end{split}
    \end{equation}
    \begin{equation}
    \label{perturbation3}
    \begin{split}
    \dot{\delta}_z=2\sqrt{2}\tilde{g}\delta_y-2\gamma_x (1-3s_z)\delta_z
    \end{split}
    \end{equation}
\end{subequations}
\begin{figure}
\includegraphics[width=0.8\linewidth]{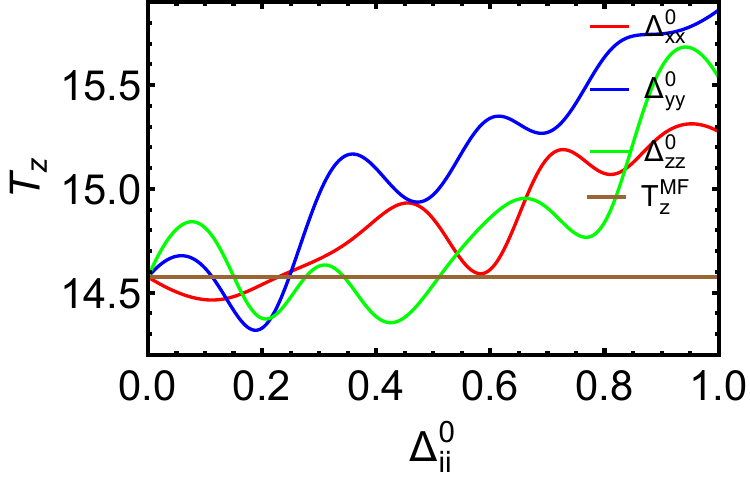}
\caption[short description]{Effect of activation of variance ($\Delta^0_{ii}$) on longitudinal relaxation time ($T_{z}$) where $\Delta^0_{xx}$ (\red), $\Delta^0_{yy}$ (\blue), $\Delta^0_{zz}$ (\green) and $T^{\text{MF}}_{z}$ (\brown) are denoted by their respective colors.
}
\label{longitudinal_relax}
\end{figure}
\begin{figure}
\includegraphics[width=0.8\linewidth]{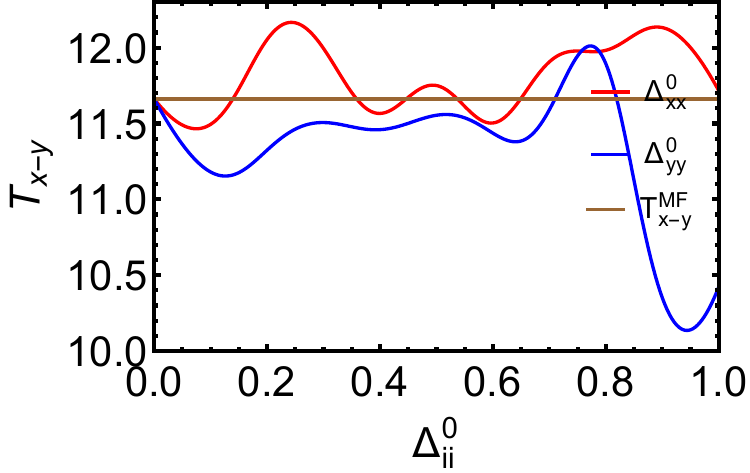}
\caption[short description]{Effect of activation of variance ($\Delta^0_{ii}$) on transverse relaxation time ($T_{x-y}$) where $\Delta^0_{xx}$ (\red), $\Delta^0_{yy}$ (\blue) and $T^{\text{MF}}_{x-y}$ (\brown) are denoted by their respective colors. Effect of $\Delta^0_{zz}$ growing on $T_{xy}$ not include here due to rapid fluctuation of $s_x$, and $s_y$.
}
\label{coherence_relaxation_time}
\end{figure}
In sec. \ref{just definition} we discuss how the longitudinal ($T_z$) and transverse ($T_{x-y}$) relaxation times of $s_i$ vary with a varying $\Delta^0_{ii}$, $i\in x,y,z$.

\subsubsection{ Longituidinal and transverse relaxation times}{\label{just definition}}

In \ref{sxyzBBR}, we have demonstrated that all Bloch vector components relax more slowly compared to their respective mean-field (MF) counterparts, if correlations and fluctuations are present in all three directions. Next, we focus on the effects of individual variances 
$\Delta^0_{ii}$  ($i\in x,y,z$). We report the effects of only the variances and not the covariances, because, as shown in \ref{sxyzBBR}, any $\Delta_{ij}$ can be expressed as $\sqrt{\Delta_{ii}\Delta_{jj}}$  ($i,j \in x,y,z$). 

 Among the relaxation times, the most significant is the longitudinal relaxation time $T_z$ that sets the scale for the population imbalance between the atomic and molecular modes to approach its equilibrium value. We observe that this $T_z$ increases with the activation of each $\Delta^0_{ii}$ as shown in Fig. (\ref{longitudinal_relax}).

 In Eq. (\ref{relaxation rate}), the decay rates of the components $s_x$, $s_y$ differ, so the decay rate in the anisotropic transverse plane can be characterized by the harmonic mean \cite{P61,P211}: \begin{equation} \label{harmonic mean} T_{x-y}=\frac{2T_xT_y}{T_x+T_y}
\end{equation}
Greater correlations lead to more coherent oscillations in the equatorial plane, i.e., $s_x$ and $s_y$ persist longer, resulting in an increased coherence time. This is shown in Fig. (\ref{coherence_relaxation_time}): we see that when $\Delta^0_{xx}$ increases, $T_{x-y}$ also increases. In contrast, an increase in $\Delta^0_{yy}$ reduces $T_{x-y}$. We could not include $T_{x-y}$ vs. $\Delta^0_{zz}$ curve in this plot due to rapid fluctuations of $s_x$, and $s_y$.\\
In the next section (Sec. \ref{long-trans}), we try to justify these results and put some physical insight behind why depending on the correlations present, the longitudinal and transverse relaxation times either get prolonged or shortened.
 
\section{Coherence dominated Relaxation time}\label{long-trans}

\begin{figure}
\centering
\begin{subfigure}{0.45\linewidth}
    \centering
    \includegraphics[width=\linewidth]{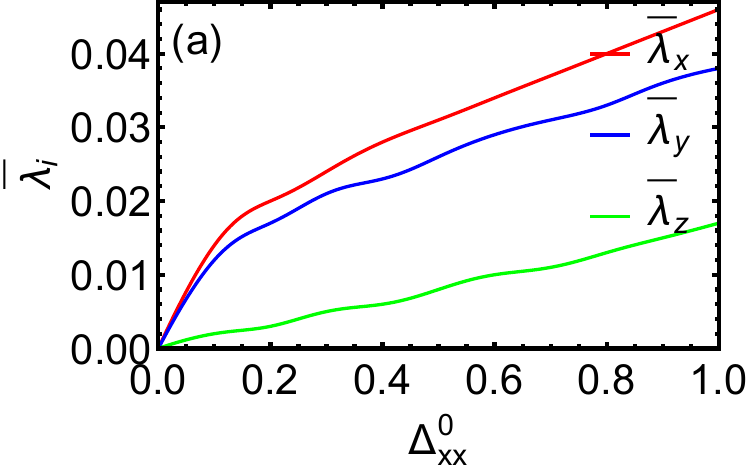}
    \phantomcaption 
    \label{noise_a}
\end{subfigure}
\begin{subfigure}{0.45\linewidth}
    \centering
    \includegraphics[width=\linewidth]{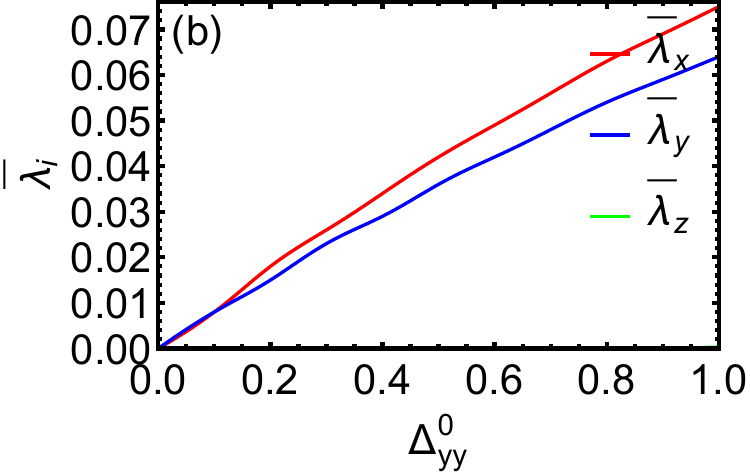}
    \phantomcaption 
    \label{noise_b}
\end{subfigure}
\begin{subfigure}{0.45\linewidth}
    \centering
    \includegraphics[width=\linewidth]{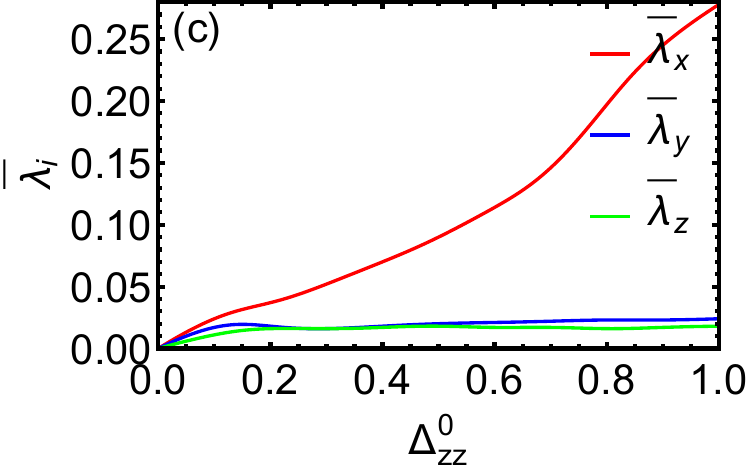}
    \phantomcaption 
    \label{noise_c}
\end{subfigure}
\caption[Short description]{time-averaged structural or elliptic noise  ($\bar{\lambda}_i$) vs. variances ($\Delta^0_{ii}$) for $\bar{\lambda}_x$ (\red), $\bar{\lambda}_y$ (\blue), and $\bar{\lambda}_z$ (\green) along $s_x$, $s_y$, and $s_z$ axis when   
(a) $\Delta^0_{xx}$,  
(b) $\Delta^0_{yy}$, and   
(c) $\Delta^0_{zz} $
 are activated.}
\label{elliptic noise}
\end{figure}

In this section, we try to provide physical arguments to explain the results obtained in Sec. \ref {Relaxation dynamics thermodynamic quantum}, especially in the presence of fluctuations and correlations. To achieve this, we take time-averages of relevant physical quantities. For this, we take the time average of any operator $\hat{A}$, such as $\langle \hat{A} \rangle_t = \bar{A}$, over the time (in our unit $t=0$ to $t=50$) during which the dynamics persists. Now, temporal averages best describe the system either under equilibrium configurations, or under non-equilibrium steady-state conditions. We understand that our system belongs to neither of these two classes and deals with a transient phenomenon (relaxation dynamics) instead. The reason we still define time averages and use them in physical interpretations is that we simply wish to capture the dominant physical traits as the system journeys back to equilibrium, and a time-averaged description works best for that purpose. Since all calculations have been performed considering weak noise,
the system can be treated in the linear response regime. Thus, the response of the system to the small noise is still governed by equilibrium fluctuations. However, nowhere in this section has the time average been equated with the ensemble average, so the question of ergodicity (or the breaking thereof) does not even arise. The time-averages that we talk about are simply indicators of the gross overall behaviors of the atoms and molecules and their approach to a steady state. \\
In Sec. \ref{squeezed}, we described the effect of elliptic noise as a squeezed state, which enhances coherence.
\subsection{Elliptic noise: Squeezed state}{\label{squeezed}}

One of the key features that emerges is that in the presence of fluctuations about the equilibrium values of the Bloch vectors, the relaxation time typically increases, especially for $s_z$. It is sort of counterintuitive because the fluctuations could help the system relax faster by facilitating transitions between different configurations. To justify our findings, we present the time-averaged structure matrix, or the real-symmetric covariance matrix,  defined as
\begin{equation}
\bar{\Delta}_{ij}= \begin{pmatrix}
\bar\Delta_{xx} & \bar\Delta_{xy} & \bar\Delta_{xz}\\
\bar\Delta_{xy} & \bar\Delta_{yy} & \bar\Delta_{yz}\\
\bar\Delta_{xz} & \bar\Delta_{yz} & \bar\Delta_{zz}
\end{pmatrix}
\end{equation}

whose eigenvalues represent the variances along the principal axes. When two of the eigenvalues of the covariance matrix are nearly equal and the third differs significantly, the noise distribution assumes an ellipsoidal shape \cite{kitagawa1993squeezed,wiener}, indicating structured fluctuations \cite{P61}, where one quadrature is expanded while the other is reduced. Such anisotropic noise leads to the formation of squeezed states. It is this structured noise that enhances the robustness of the non-equilibrium system by suppressing specific decoherence channels, thereby increasing the relaxation time or prolonging coherence. In our system, increasing $\Delta^0_{xx}$ and $\Delta^0_{yy}$ causes $\bar{\lambda}_x$ and $\bar{\lambda}y$ to become nearly equal, while $\bar{\lambda}z$ remains significantly smaller as shown in Figs. (\ref{noise_a}) and (\ref{noise_b}). When $\Delta^0_{zz}$ increases, $\bar{\lambda}_y$ and $\bar{\lambda}_z$ become nearly equal, and $\bar{\lambda}_x$ grows substantially, as illustrated in Fig. (\ref{noise_c}).  In both cases, the noise exhibits an elliptic structure, suggesting that the system's dynamics becomes increasingly constrained, and therefore, relaxation occurs more slowly.\\

\begin{figure}
\centering
    \includegraphics[width=\linewidth]{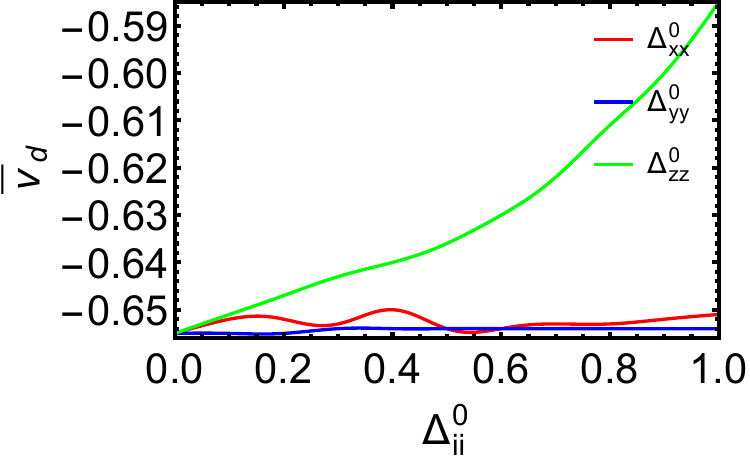}
\caption[short description]{variance ($\Delta^0_{ii}$) motivated time-averaged drift speed ($\bar v_d$) 
where the following correlations are represented by their respective colors for :    
$\Delta^0_{xx}$ (\red),  
$\Delta^0_{yy}$ (\blue), and 
$\Delta^0_{zz}$ (\green)
}
\label{drift}
\end{figure}

\begin{figure}
\centering
    \includegraphics[width=\linewidth]{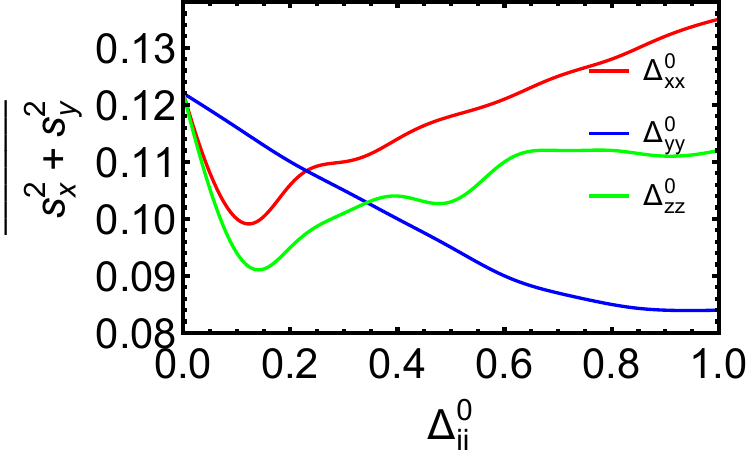}
\caption[short description]{Time-averaged fringe-visibility ($\overline{s_x^2+s_y^2}$) dominated by fluctuations where the following variances ($\Delta^0_{ii}$) are represented by their respective colors for :    
$\Delta^0_{xx}$ (\red),  
$\Delta^0_{yy}$ (\blue), and   
$\Delta^0_{zz}$ (\green)
}
\label{fringe}
\end{figure}

\begin{figure}
    \centering
    \includegraphics[width=\linewidth]{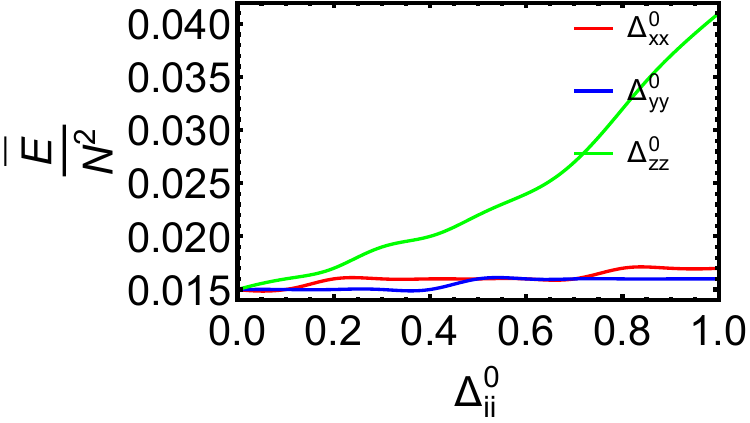}
\caption[short description]{time-averaged Phase-entanglement  ($\bar E$) controlled by $\Delta^0_{ii}$ where the following correlations are represented by their respective colors for :    
$\Delta^0_{xx}$ (\red),  
$\Delta^0_{yy}$ (\blue), and 
$\Delta^0_{zz}$ (\green)
}
\label{entropy}
\end{figure}
\begin{figure}
    \centering
    \includegraphics[width=\linewidth]{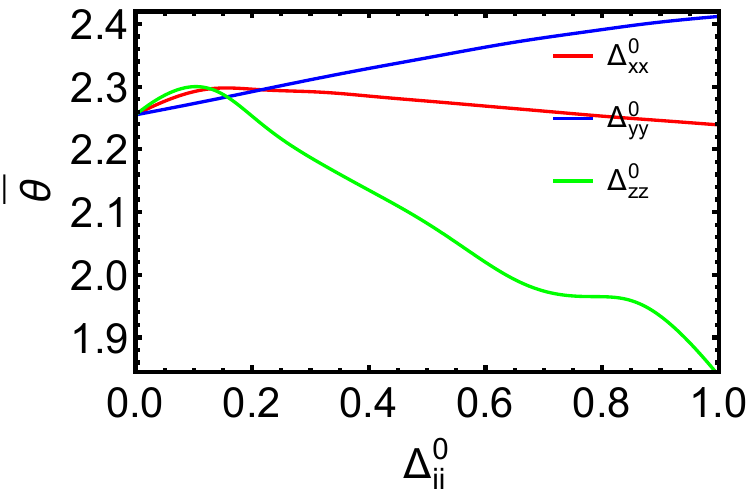}
\caption[short description]{time-averaged polar angle ($\overline{\theta}$) controlled by $\Delta^0_{ii}$ where the following variances are represented by their respective colors for :    
$\Delta^0_{xx}$ (\red),  
$\Delta^0_{yy}$ (\blue), and 
$\Delta^0_{zz}$ (\green)
 
}
\label{theta}
\end{figure}

\subsection{Other time-averaged quantities as indicators of coherence }\label{intution}

So far, we have observed that the longitudinal relaxation time increases under the influence of any type of fluctuations. On the contrary, the transverse relaxation time increases with an increasing $\Delta^0_{xx}$ or $\Delta^0_{zz}$, and decreases with an increasing $\Delta^0_{yy}$. Now, we study the effect of correlations on a few relevant physical quantities, and try to tally the findings with the trends that arise in the relaxation process. Three such quantities that we address here are (i) drift speed, (ii) fringe visibility, (iii) EPR entanglement, and (iv) the polar angle of the Bloch vector. 

\subsubsection{Drift speed}
We notice that the dtift speed decreases \cite{khodorkovsky2009decoherence} with an increase in structured quantum noise \cite{P10} as shown in Fig. (\ref{longitudinal_relax}). Note that the drift speed, defined as $\dot{s}_z$, represents the tunneling speed between the molecular and atomic states.

\subsubsection{Fringe visibility}

This quantity can be captured in time-of-flight interference experiments \cite{flight}. We have studied the fringe-visibility of our system using density matrix formalism. Here, the density matrix can be expressed as \cite{wiener,P510}: \begin{equation} \hat{\rho}=\frac{I+\sigma_j s_j}{2},\quad \text{where} \quad j\in {x,y,z} \label{density matrix} \end{equation}
Here, $\sigma_j$ are $(2\cross 2)$ Pauli matrices.
In matrix form, Eq. (\ref{density matrix}) becomes: 
    \begin{equation}\label{eigen value} \hat{\rho}=\frac{1}{2}\begin{pmatrix}
1+s_z & s_x-is_y\\ s_x+is_y & 1-s_z \end{pmatrix}
\end{equation}
the magnitude of this off-diagonal element, $s_\parallel$ is called fringe-visibility.
We calculate the the evolution of the time-averaged transverse component ($\bar{s}_\parallel =\overline{s_x^2+s_y^2}$) \cite{hao2011dynamics,chuchem2010quantum} with increasing initial variances. We observe that $\overline{s_x^2+s_y^2}$ increases with an increasing $\Delta^0_{xx}$ or $\Delta^0_{zz}$, and decreases with a decreasing $\Delta^0_{yy}$.  Since coherence corresponds to the magnitude of the off-diagonal elements of the density matrix, the dephasing mechanism can be identified when the average projection length ($S_\parallel$) shrinks. So clearly, a growth in the fringe visibility implies a larger relaxation time, and a decay in the visibility indicates a faster relaxation process, as evident from Figs. (\ref{coherence_relaxation_time}) and (\ref{fringe}).

Actually, fringe visibility is the amplitude of population oscillation.

.\\  

\subsubsection{EPR entanglement}
The  Einstein-Podolski-Rosen (EPR) entanglement or phase entanglement measurement \cite{pudlik2013dynamics} $E$ is another key element to study the coherence of the system, which is defined as,
 \begin{subequations}
  \begin{equation}
    E=\langle{\hat{L}_+\hat{L}_-}\rangle-\langle 2\hat{n}_b\rangle\langle \hat{n}_a\rangle
\end{equation}
\text{which recasts as large N limit}
\begin{equation}
\frac{E}{N^2}\approx\frac{2s^2_z+\Delta_{zz}}{8}
\end{equation}   
 \end{subequations}

where, $\hat{L}_+=\hat{a}^\dagger\hat{a}^\dagger\hat{b}$, and $\hat{L}_-=\hat{b}^\dagger \hat{a}\hat{a}$. From Fig. (\ref{entropy}), we observe that $E$ increases when either $\Delta^0_{xx}$ or $\Delta^0_{zz}$ increases. On the other hand, if $\Delta^0_{yy}$ increases, then $E$ slightly decreases: again, consistent with the corresponding relaxation processes.\\ 

\subsubsection{Variation of the polar angle }

In Sec. \ref{sxyzBBR} and Sec. \ref{intution}, it has been discussed how introduction of the variances $\Delta_{xx}$ and $\Delta_{zz}$ influence the system in a similar way, while the impact of $\Delta_{yy}$ is exactly opposite. This can be justified by studying the polar angle ($\theta$) between $s_z$ axis and $\mathbf{s}$ in Fig. (\ref{bloch sphere}). This $\theta$ can be defined as,
\begin{equation}
    \theta=\cos^{-1}\frac{s_z}{\mid\mathbf{s}\mid}
\end{equation}
Its variation with each $\Delta^0_{ii}$ is shown in Fig. (\ref{theta}). We find that the time-averaged polar angle lies in 2nd quadrant, i.e., $\frac{\pi}{2}<\bar{\theta}<\pi$, which means, if $\bar{\theta}$ decreases then the precession cone becomes  wider like a classical top motion \cite{P-304}. If the radius of precession cone ($s_\parallel$) increases then coherence also increases because, $s_\parallel$ is nothing but the square root of the Fringe Visibility.  If $\theta$ moves from $\pi$ to $\frac{\pi}{2}$, then $s_x\approx\cos\theta$ remains while  $s_y\approx\sin\theta$ becomes negative. Thus, in our regime of interest,  $s_x$, and $s_y$ bear opposite signs.

In the  equatorial plane ($s_z=0$), $s_\parallel$ reaches its maximum because \cite{bloch4}
\begin{equation}
\label{opposite}
    s^2_x+s^2_y=\frac{(1+s_z)(1-s_z)^2}{2}
\end{equation}
If we add small fluctuation on eac $s_i$ as $\delta s_i$ $(i\in x,y,z)$ then Eq. (\ref{opposite}) becomes
\begin{equation}
    s_x\delta s_x+s_y\delta s_y=-\frac{(1-s_z)(1+3s_z)\delta s_z}{2}
\label{fluctuation}
\end{equation}
Since $\bar\theta$ remains in 2nd quadrant, $s_x < 0$ and $s_y >0$ in Eq. (\ref{fluctuation}). Thus, $\delta s_x$ and $\delta s_z$ carry same signs, and that of $\delta s_y$ is the opposite.  As a result, the impact of $\Delta_{xx}$ and $\Delta_{zz}$ are closer, while that of $\Delta_{yy}$ is the opposite. Therefore, in Figs. (\ref{coherence_relaxation_time}), (\ref{fringe}), (\ref{entropy}), and (\ref{theta}) an activation of $\Delta_{yy}$ works in opposite of an activation of $\Delta_{xx}$ and $\Delta_{zz}$.

\section{Summary and Conclusion}\label{conclusion}

In this work, we have studied a system of bosonic atoms coupled via a Feshbach resonance, capable of forming  bosonic dimers. We have corrupted both the Feshbach detuning and the Feshbach coupling  by Gaussian white noise, and studied the corresponding relaxation dynamics. We have described the system in a Bloch sphere representation. where ($\hat{L}_z$) denotes the population difference between atomic and molecular Bose-Einstein condensates (BECs), and  $\hat{L}_x$ and $\hat{L}_y$ represent the real and imaginary parts of the coherences, respectively.

We adopted two methods for studying the dynamics. In the mean-field (MF) approach, only the mean values are considered: $s_i = \langle \hat{L}_i \rangle$. In contrast, the Bogoliubov Backreaction (BBR) method accounts for both the mean and the second moments i.e., variances ($\Delta^0_{ii}$) and covariances ($\Delta^0_{ij}$).

Since the system contains particle-conserving noise (i.e., $\dot{N} = 0$), the relaxation times of both polarization ($s_\perp$) and coherence ($s_\parallel$) are characterized by the longitudinal ($T_{z}$) and transverse ($T_{x-y}$) relaxation times, respectively. Here, we observe that $T_z$ always increases under the influence  of any rising $\Delta^0_{ii}$. However, $T_{x-y}$ increases when $\Delta^0_{xx}$ and $\Delta^0_{zz}$ increase, whereas an increase in $\Delta^0_{yy}$ drives the system towards a smaller $T_{x-y}$. We found a connection of this result with the behaviour of several related physical quantities : 
(i)  drift speed ($v_d$)  (ii) fringe visibility ($s_x^2 + s_y^2$), (iii) EPR entanglement ($E$), and (iv) polar angle ($\theta$). \\

In the present manuscript, the influence of the variances $\Delta^0_{ii}$ and covariances $\Delta^0_{ij}$ of the Bloch vector on the system evolution is reported only.  However, Feshbach detuning and initial polarization also play important roles in controlling the system dynamics. we intend to investigate these two aspects in future. Moreover, another interesting phenomenon observed in certain quantum system is that a certain amount of noise \cite{witthaut2008dissipation,witthaut2009dissipation,jacobo2010effects} can enhance coherence a phenomenon known as coherence resonance. A study of this coherence resonance for the atomic-molecular condensates is also on our cards.
 
\bibliography{bibi.bib}

\end{document}